\documentclass{pasj01}
\draft
\Received{$\langle$reception date$\rangle$}
\Accepted{$\langle$acception date$\rangle$}
\Published{$\langle$publication date$\rangle$}
\begin{document}

\title{BM UMa: a middle shallow contact binary at pre-transition stage of evolution from W-type to A-type}

\author{T. Sarotsakulchai\altaffilmark{1}, B. Soonthornthum\altaffilmark{1}, S. Poshyachinda\altaffilmark{1}, C. Buisset\altaffilmark{1}, T. Lépine\altaffilmark{2,3}, A. Prasit\altaffilmark{1}}

\email{tavijarus@narit.or.th}

\altaffiltext{1}{National Astronomical Research Institute of Thailand, Donkaew, Maerim, Chiangmai 50180, Thailand}
\altaffiltext{2}{Univ-Lyon, Laboratoire Hubert Curien, UMR CNRS 5516, 18 rue Benoît Lauras, 42000 Saint-Etienne,
France}
\altaffiltext{3}{Institut d'Optique Graduate School, 18 rue Benoît Lauras, 42000 Saint-Etienne, France}

\maketitle

\begin{abstract}

In this study, all unpublished time series photometric data of BM UMa ($q \sim$ 2.0, P = 0.2712\,d) from available archives were re-investigated together with new data taken from the TNT-2.4m of the Thai National Observatory (TNO). Based on period analysis, there is a short-term variation superimposed on the long-term period decrease. The trend of period change can be fitted with a downward parabolic curve indicating a period decrease at a rate of $\mathrm{d}P/\mathrm{d}t = -3.36(\pm 0.02)\times10^{-8}$ d $\textrm{yr}^{-1}$. This long-term period decrease can be explained by mass transfer from the more massive component ($M_2 \sim 0.79 M_{\odot}$) to the less massive one ($M_1 \sim 0.39 M_{\odot}$), combination with AML. For photometric study, we found that the binary consists of K0\,V stars and at the middle shallow contact phase with evolution of fill-out factor from 8.8\,\% (in 2007) to 23.2\,\% (in 2020). Those results suggest that the binary is at pre-transition stage of evolution from W-type to A-type, agreeing to the results of statistical study of W-type contact binaries. The mass of $M_2$ will be decreased close to or below $M_1$ and the mass ratio will be decreased ($q < 1.0$). By this way, the binary will evolve into A-type as a deeper normal over-contact system with period increase. Finally the binary will end as a merger or a rapid-rotating single star when the mass ratio meet the critical value ($q < 0.094$), as well as produce a red nova.

\end{abstract}

\section{Introduction}

EW-type Eclipsing binaries are commonly composed of main-sequence F, G and K-type stars (Rucinski 2007; Qian et al. 2017). Most of them are solar-type dwarfs with strong stellar activities which can be seen from their light-curve asymmetries and spectral line profile (chromospheric activity e.g., strong Ca II H and K emission). Contact binaries with K-type dwarfs are known for decades (see Bradstreet 1985) and to date, there are many contact binaries with K-type stars that were studied (e.g., Yang et al. 2008, 2009a, 2011; Liao et al 2012; Zhu et al. 2014; Liu et al. 2014a, 2014b, 2015, 2020; Martignoni et al. 2016; Xia et al. 2018; Zhou \& Soonthornthum 2019a, 2020). Most of them are shallow W-subtype contact binaries with light curve asymmetries. Recent study by Qian et al. (2017), pointed out that the period distribution of EW-type binaries which observed by the Large Sky Area Multi-Object Fiber Spectroscopic Telescope (LAMOST; Cui et al. 2012) shows a peak close to 0.29\,d. Similar study by Rucinski (2007), reported the period study from the All-Sky Automated Survey (ASAS; Pojmanski 1997) at a peak about 0.27\,d, while Jiang et al. (2014) reported a shorter period at a peak of 0.25\,d from the study of Kepler data archive. In addition, the study by Qian et al. (2017) also suggested that the peak of temperature distribution is about 5700 K which corresponds to the temperature of a G3-type or solar-type star and the peak of distribution of gravitational acceleration log (g) is close to 4.16. They noted that those results are in agreement with the idea that contact binaries are formed from detached binaries via the Case A mass transfer or/and shrinking orbits via angular momentum loss (AML) due to magnetic braking (Bradstreet \& Guinan 1994).

BM UMa is a W-subtype of contact binary with a period of about 0.2712\,d and magnitude about 14 in V-band. It was firstly discovered by Hoffmeister (1963) and correctly identified as an eclipsing W UMa type binary by Shugarov (1975). The recent study by Samec et al. (1992; 1995), who published its first CCD observations, new period analysis and ephemeris, indicated that its light curve asymmetries would be migration by a hot spot in the neck of the secondary component where both components were classified as early K-type stars. They noted that this may be attributed to fluid dynamics of mass in transit rather than to magnetic activity. Their period study showed that the period of the binary are decreasing, which is in agreement with the analysis results from Yang et al. (2009), who found a cyclic oscillation superimposed on a secular period decrease in the time interval over 77 years. Yang et al. (2009) also suggested that the long-term period decrease may be due to mass transfer between the component and/or mass and angular-momentum loss (AML) from the system, while the cyclic change may be due to a light travel-time effect (LTTE) from an unseen third body with period of 30.8 yrs. However, no asymmetry in the light curves or the O'Connell effect was reported from Yang et al. (2009), just only a large scatter or mean errors due to faint star observations (e.g., 0.065 mag for V-band and 0.077 mag for R-band) was noted in the paper. Later, Virnina et al. (2010) published the new results of V-band CCD observations and obtained full light-curves for BM UMa. They showed that the light curves were symmetric without the O'Connell effect and they also did not confirm the presence of spots on the components. As their light curve investigation indicated that the mass ratio of the system was 0.54 where degree of contact about 10.7, with temperature of 4700 K and 4510 K for the primary and the secondary component, respectively.

In this paper, we will re-investigate the binary system BM UMa with new photometric light curves in VRI bands from our observations via the 2 meter class telescope together with an analysis of long-term photometric time series from available surveys which are unpublished before. The new results from period study are also presented together with new aspect from long-term photometric time series. The magnetic activities and the short-term period change are also investigated and compared to the other K-type contact binaries.

\section{New photometric observations and light curves}

In April 2, 2020, we performed the observations of BM UMa ($\alpha_{2000}$ = $11^{\textrm{h}}11^{\textrm{m}}20^{\textrm{s}}.5$ and $\delta_{2000}$ = $+46^\circ25^{\prime}47^{\prime\prime}.3$) by using the 2.4-m Thai National Telescope (TNT) at the Thai National Observatory (TNO), Chiangmai, Thailand. For data acquisitions, the instrument system e.g., the TNT focal reducer and the ARC 4k camera were used. For more information about the telescope and the instrument system, see Prasit et al. (2019). During the observations, VRI filters were employed. All images from the observations were processed by using a standard procedure in IRAF\footnote{The Image Reduction and Analysis Facility (IRAF), http://iraf.noao.edu.}. The comparison ($\alpha_{2000}$ = $11^{\textrm{h}}11^{\textrm{m}}22^{\textrm{s}}.4$ and $\delta_{2000}$ = $+46^\circ21^{\prime}51^{\prime\prime}.7$) and the check stars ($\alpha_{2000}$ = $11^{\textrm{h}}11^{\textrm{m}}47^{\textrm{s}}.5$ and $\delta_{2000}$ = $+46^\circ24^{\prime}21^{\prime\prime}.5$) are used for differential photometry. The full light curves of BM UMa in VRI bands are plotted together and compared to the check star as displayed in Fig 1. The figure shows that all bands of light curves are symmetric without the O'Connell effect, where the light curves (VRI) of the check star are mostly stable except the observations in the early night with bad seeing affecting on the data and showing small scatters.

\begin{figure}
\begin{center}
\includegraphics[angle=0,scale=0.4]{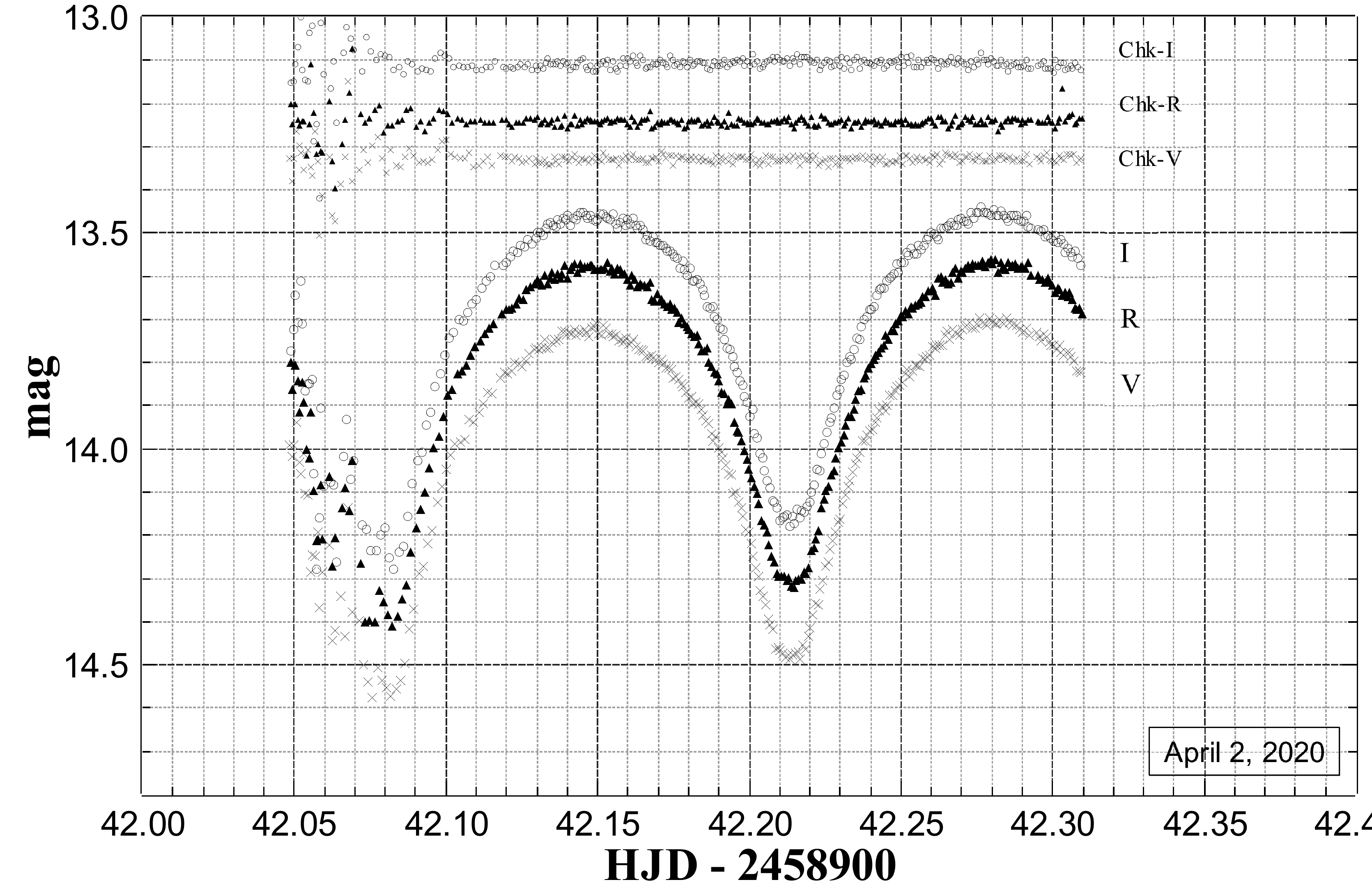}
\caption{The multi-color CCD light curves of BM UMa in V, R and I bands were obtained with the TNT-2.4m in April 2020. The differential magnitude between the comparison and the check stars (C-Ch) in VRI bands are also displayed in the figure without the significant variation except in the early night with bad seeing.}
\end{center}
\end{figure}

\section{Orbital period study}

The period changes of BM UMa were studied by many authors e.g., Samec \& Kreidl (1992); Samec et al. (1995); Yang et al. (2009) and Virnina et al. (2010), but it has been neglected for years since 2010. To re-investigate the period variation, we collected all available observed eclipse times (times of minimum light: ToM) including data from photographic (pg), visual (vis), photoelectric (pe) and ccd methods which reported in the literature as references shown in the last column of Table 2. The orbital period variation of BM UMa can be analyzed from eclipse timing differences by plotting the $O-C$ diagram, where $O$ refers to the observed eclipse times and $C$ refers to the computed ones which predicted by the reference epoch (Kreiner 2004). All $O-C$ values were determined by using the linear ephemeris in Eq. (1), where the corresponding values are listed in Table 1 and the result is plotted in Fig. 2. 

\begin{equation}
Min.I (HJD) = 2444292.3496+0^{d}.2712208E
\end{equation}

\begin{figure}
\begin{center}
\includegraphics[angle=0,scale=0.4]{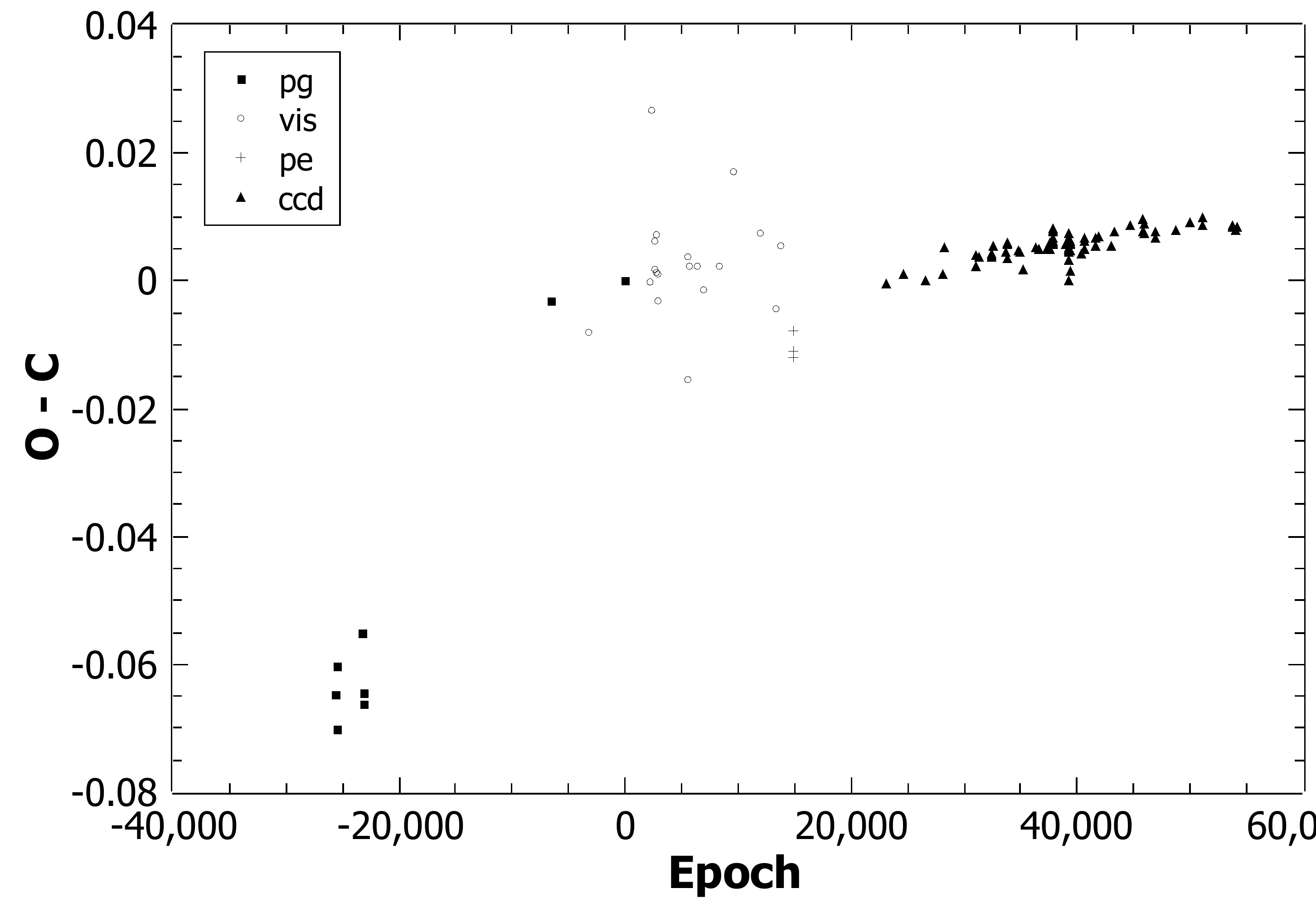}
\caption{The $O-C$ diagram is constructed by using the linear ephemeris in Eq. (1) where data from all observation methods were compiled. The filled-square marks refer to the $O-C$ data from photographic (pg) observations, while the open circles refer to the visual (vis) ones. The plus (+) and the filled-triangle marks refer to the photoelectric (pe) and the ccd methods, respectively.}
\end{center}
\end{figure}

The plot in Fig. 2 shows that the two data from pg method by Hoffmann (1981), Shugarov (1975) and some data from visual method are not following to the general trend of the $O-C$ curve. To get the best fit, we decide to remove those $O-C$ data obtained from visual method and mainly use data from pe and ccd methods, as well as the early 6 values from pg method by Busch et al. (1966) and 7 additional ToM data determined from the latest AAVSO\footnote{the American Association of Variable Star Observers: AAVSO; https://www.aavso.org} observations 2016 - 2020. By using a least-squares method the curve fitting and the residuals are plotted in Fig. 3, while the improved ephemeris is given in Eq. (2), respectively.

\begin{equation}
\begin{array}{ll}
Min.I(HJD) =  2444292.3245(\pm 0.0003) +0.27122209(\pm 0.00000001)E 
\\ -[12.50(\pm 0.27)\times 10^{-12}]E^2
\end{array}
\end{equation}

According to Fig. 3, it can be seen that the general trend of the variation can be fitted with a quadratic ephemeris in Eq. (2) where 
the quadratic term in Eq. (2) reveals a continuous period decrease at a rate of $\mathrm{d}P/\mathrm{d}t = -3.36(\pm 0.02)\times10^{-8}$ d $\textrm{yr}^{-1}$. After the long-term period change is subtracted from the $O-C$ diagram, the scatters are still presented shown in the lower panel of Fig. 3. The residuals of Eq. (2) plotted in the lower panel of Fig. 3 suggest that the changes of the orbital period of BM UMa are complex due to a small-amplitude variation and scatters. This may indicate that only an downward parabolic curve cannot fit the $O-C$ data very well. However, to get a better fit for the trend of the $O-C$ data and to prove the existence of short-term variations superimposed on the long-term period decrease in BM UMa, long-term observations and new eclipse times are needed in the future. 

\begin{figure}
\begin{center}
\includegraphics[angle=0,scale=0.7]{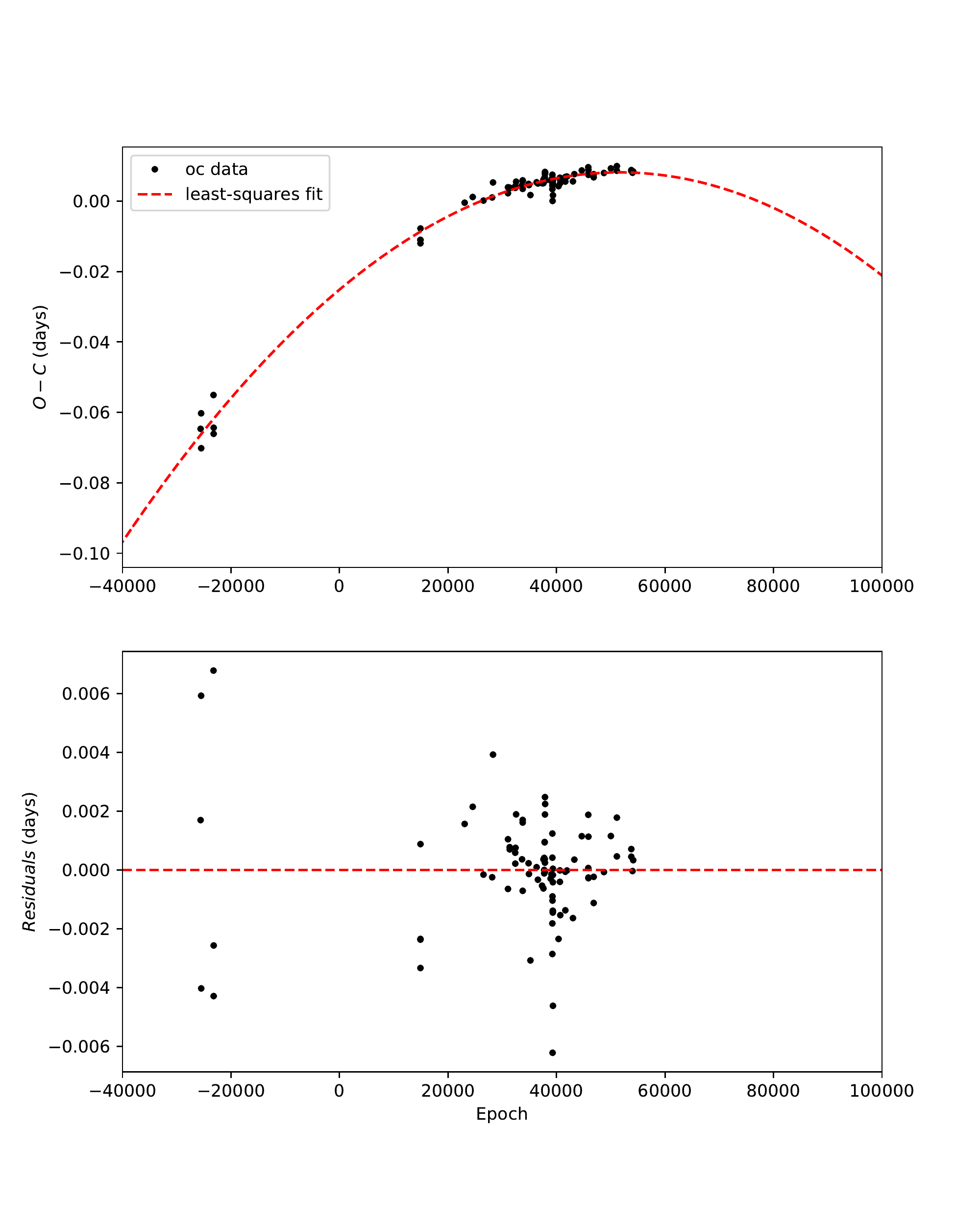}
\caption{The curve fitting (upper panel) for the $O-C$ data by using the quadratic ephemeris in Eq. (2) is plotted as dashed line with an downward parabolic curve. The least-squares fit shows that the long-term period change is decreasing and the peak is close to the recent observations in 2020 at cycle E = 50,000. The lower panel shows the result after subtracting the quadratic term from the $O-C$ data displayed as the residuals from Eq. (2). The residuals show the variations of scatter which may indicate the existence of short-term period change which superimposed on the long-term period decrease.}
\end{center}
\end{figure}

\begin{table}
\scriptsize
\caption{Times of minimum light of BM UMa}
\begin{center}
\begin{tabular}{lcrrccc}\hline\hline
HJD(2400000+) & Error(days) & E & $(O-C)_1$ & Method & Min & Ref. \\
\hline
37348.490	&		&	-25602.0	&	-0.06468	&	pg	&	I	&	Busch et al. (1966)	\\
37374.396	&		&	-25506.5	&	-0.06026	&	pg	&	II	&	Busch et al. (1966)	\\
37378.590	&		&	-25491.0	&	-0.07019	&	pg	&	I	&	Busch et al. (1966)	\\
37992.649	&		&	-23227.0	&	-0.05508	&	pg	&	I	&	Busch et al. (1966)	\\
38000.639	&		&	-23197.5	&	-0.06609	&	pg	&	II	&	Busch et al. (1966)	\\
38001.590	&		&	-23194.0	&	-0.06436	&	pg	&	I	&	Busch et al. (1966)	\\
42511.375	&		&	-6566.5	&	-0.00322	&	pg	&	II	&	Shugarov (1975)	\\
43433.521	&		&	-3166.5	&	-0.00794	&	vis	&	II	&	http://var2.astro.cz/ocgate/	\\
44292.3496	&	0.0008	&	0.0	&	0	&	pg	&	I	&	Hoffmann (1981)	\\
44292.4853	&		&	0.5	&	0.00009	&	pg	&	II	&	Hoffmann (1981)	\\
44885.645	&		&	2187.5	&	-0.0001	&	vis	&	II	&	http://var2.astro.cz/ocgate/	\\
44910.624	&		&	2279.5	&	0.02659	&	vis	&	II	&	http://var2.astro.cz/ocgate/	\\
44987.626	&		&	2563.5	&	0.00188	&	vis	&	II	&	http://var2.astro.cz/ocgate/	\\
44994.682	&		&	2589.5	&	0.00614	&	vis	&	II	&	http://var2.astro.cz/ocgate/	\\
45053.538	&		&	2806.5	&	0.00722	&	vis	&	II	&	http://var2.astro.cz/ocgate/	\\
45056.380	&		&	2817.0	&	0.00141	&	vis	&	I	&	http://var2.astro.cz/ocgate/	\\
45061.393	&		&	2835.5	&	-0.00318	&	vis	&	II	&	http://var2.astro.cz/ocgate/	\\
45075.365	&		&	2887.0	&	0.00095	&	vis	&	I	&	http://var2.astro.cz/ocgate/	\\
45790.577	&		&	5524.0	&	0.0037	&	vis	&	I	&	http://var2.astro.cz/ocgate/	\\
45813.476	&		&	5608.5	&	-0.01546	&	vis	&	II	&	http://var2.astro.cz/ocgate/	\\
45830.445	&		&	5671.0	&	0.00224	&	vis	&	I	&	http://var2.astro.cz/ocgate/	\\
46039.692	&		&	6442.5	&	0.0024	&	vis	&	II	&	http://var2.astro.cz/ocgate/	\\
46173.400	&		&	6935.5	&	-0.00146	&	vis	&	II	&	http://var2.astro.cz/ocgate/	\\
46561.385	&		&	8366.0	&	0.00219	&	vis	&	I	&	http://var2.astro.cz/ocgate/	\\
46907.342	&		&	9641.5	&	0.01706	&	vis	&	II	&	http://var2.astro.cz/ocgate/	\\
47524.631	&		&	11917.5	&	0.00752	&	vis	&	II	&	http://var2.astro.cz/ocgate/	\\
47927.382	&		&	13402.5	&	-0.00437	&	vis	&	II	&	http://var2.astro.cz/ocgate/	\\
48039.406	&		&	13815.5	&	0.00544	&	vis	&	II	&	http://var2.astro.cz/ocgate/	\\
48338.9529	&	0.0004	&	14920.0	&	-0.01104	&	pe	&	I	&	Samec \& Kreidl (1992)	\\
48339.7656	&	0.0004	&	14923.0	&	-0.012	&	pe	&	I	&	Samec \& Kreidl (1992)	\\
48339.9022	&	0.0006	&	14923.5	&	-0.01101	&	pe	&	II	&	Samec \& Kreidl (1992)	\\
48340.8547	&		&	14927.0	&	-0.00778	&	pe	&	I	&	Samec \& Kreidl (1992)	\\
50549.4130	&		&	23070.0	&	-0.00046	&	ccd	&	I	&	http://var2.astro.cz/ocgate/	\\
50953.3980	&		&	24559.5	&	0.00116	&	ccd	&	II	&	http://var2.astro.cz/ocgate/	\\
51490.8210	&		&	26541.0	&	0.00015	&	ccd	&	I	&	http://var2.astro.cz/ocgate/	\\
51926.5381	&		&	28147.5	&	0.00103	&	ccd	&	II	&	http://var2.astro.cz/ocgate/	\\
51967.4967	&		&	28298.5	&	0.00529	&	ccd	&	II	&	http://var2.astro.cz/ocgate/	\\
52715.3850	&		&	31056.0	&	0.00224	&	ccd	&	I	&	http://var2.astro.cz/ocgate/	\\
52715.5223	&		&	31056.5	&	0.00392	&	ccd	&	II	&	http://var2.astro.cz/ocgate/	\\
52800.4143	&	0.0001	&	31369.5	&	0.00381	&	ccd	&	II	&	Kotkova \& Wolf (2006)	\\
52801.3635	&	0.0003	&	31373.0	&	0.00374	&	ccd	&	I	&	Kotkova \& Wolf (2006)	\\
53076.6530	&	0.0003	&	32388.0	&	0.00413	&	ccd	&	I	&	Dvorak (2005)	\\
53081.3990	&	0.0001	&	32405.5	&	0.00377	&	ccd	&	II	&	Kotkova \& Wolf (2006)	\\
53092.3840	&	0.0001	&	32446.0	&	0.00432	&	ccd	&	I	&	Krajci (2005)	\\
53122.4907	&	0.0005	&	32557.0	&	0.00551	&	ccd	&	I	&	Hubscher et al. (2005)	\\
53418.7984	&	0.0001	&	33649.5	&	0.00449	&	ccd	&	II	&	Nelson (2006)	\\
53451.3462	&	0.0001	&	33769.5	&	0.00579	&	ccd	&	II	&	Hubscher et al. (2005)	\\
53451.4819	&	0.0002	&	33770.0	&	0.00588	&	ccd	&	I	&	Hubscher et al. (2005)	\\
53451.6151	&	0.0007	&	33770.5	&	0.00347	&	ccd	&	II	&	Hubscher et al. (2005)	\\
53744.9418	&	0.0002	&	34852.0	&	0.00488	&	ccd	&	I	&	Dvorak (2008)	\\
53763.6557	&	0.0021	&	34921.0	&	0.00454	&	ccd	&	I	&	Brat et al. (2007)	\\
53837.4249	&	0.0026	&	35193.0	&	0.00172	&		&	I	&	SWASP*	\\
54145.8066	&	0.0001	&	36330.0	&	0.00534	&	ccd	&	I	&	Samolyk (2016a)	\\
54211.4417	&	0.0002	&	36572.0	&	0.00500	&	ccd	&	I	&	Dogru et al. (2007)	\\
54419.0613	&	0.0002	&	37337.5	&	0.00508	&	ccd	&	II	&	Nelson (2008)	\\
\hline
\end{tabular}
\end{center}
{\tiny Notes.} \tiny *We determined the times of light minima of BM UMa's light curves from SWASP database (https://wasp.cerit-sc.cz/form, see Fig. 4)
\end{table}

\begin{table}
\scriptsize
\caption{Times of minimum light of BM UMa (continued)}
\begin{center}
\begin{tabular}{lcrrccc}\hline\hline
HJD(2400000+) & Error(days) & E & $(O-C)_1$ & Method & Min & Ref. \\
\hline
54488.2226	&	0.0010	&	37592.5	&	0.00508	&	ccd	&	II	&	Yang et al. (2009)	\\
54488.3592	&	0.0021	&	37593.0	&	0.00607	&	ccd	&	I	&	Yang et al. (2009)	\\
54526.1946	&	0.0002	&	37732.5	&	0.00616	&	ccd	&	II	&	Yang et al. (2009)	\\
54526.3298	&	0.0003	&	37733.0	&	0.00575	&	ccd	&	I	&	Yang et al. (2009)	\\
54527.8214	&	0.0001	&	37738.5	&	0.00564	&	ccd	&	II	&	Nelson (2009)	\\
54550.0626	&	0.0003	&	37820.5	&	0.00673	&	ccd	&	II	&	Yang et al. (2009)	\\
54550.1982	&	0.0003	&	37821.0	&	0.00672	&	ccd	&	I	&	Yang et al. (2009)	\\
54554.9440	&	0.0003	&	37838.5	&	0.00616	&	ccd	&	II	&	Yang et al. (2009)	\\
54562.9449	&	0.0002	&	37868.0	&	0.00605	&	ccd	&	I	&	Yang et al. (2009)	\\
54563.0802	&	0.0002	&	37868.5	&	0.00574	&	ccd	&	II	&	Yang et al. (2009)	\\
54564.9807	&	0.0009	&	37875.5	&	0.00769	&	ccd	&	II	&	Yang et al. (2009)	\\
54565.1169	&	0.0007	&	37876.0	&	0.00828	&	ccd	&	I	&	Yang et al. (2009)	\\
54575.4231	&	0.0029	&	37914.0	&	0.00806	&		&	I	&	SWASP*	\\
54845.8280	&	0.0004	&	38911.0	&	0.00585	&	ccd	&	I	&	Deithelm (2009)	\\
54887.5962	&	0.0001	&	39065.0	&	0.00605	&	ccd	&	I	&	Samolyk (2010)	\\
54933.9733	&	0.0003	&	39236.0	&	0.00443	&	ccd	&	I	&	Virnina et al. (2010)	\\
54934.7860	&	0.0002	&	39239.0	&	0.00339	&	ccd	&	I	&	Virnina et al. (2010)	\\
54935.3325	&	0.0003	&	39241.0	&	0.00749	&	ccd	&	I	&	Hubscher (2011)	\\
54935.6029	&	0.0019	&	39242.0	&	0.00667	&	ccd	&	I	&	Hubscher (2011)	\\
54939.6699	&	0.0001	&	39257.0	&	0.00535	&	ccd	&	I	&	Virnina et al. (2010)	\\
54941.7039	&	0.0002	&	39264.5	&	0.00521	&	ccd	&	II	&	Virnina et al. (2010)	\\
54946.3095	&	0.0002	&	39281.5	&	0.00004	&	ccd	&	II	&	Samolyk (2010)	\\
54955.6715	&	0.0002	&	39316.0	&	0.00489	&	ccd	&	I	&	Virnina et al. (2010)	\\
54955.8070	&	0.0003	&	39316.5	&	0.00482	&	ccd	&	II	&	Virnina et al. (2010)	\\
54958.7915	&	0.0003	&	39327.5	&	0.00586	&	ccd	&	II	&	Virnina et al. (2010)	\\
54959.8766	&	0.0005	&	39331.5	&	0.0061	&	ccd	&	II	&	Virnina et al. (2010)	\\
54960.8246	&	0.0006	&	39335.0	&	0.00488	&	ccd	&	I	&	Virnina et al. (2010)	\\
54961.7754	&	0.0002	&	39338.5	&	0.00632	&	ccd	&	II	&	Virnina et al. (2010)	\\
54964.3473	&	0.0002	&	39348.0	&	0.00166	&	ccd	&	I	&	Samolyk (2010)	\\
55243.7073	&	0.0003	&	40378.0	&	0.00424	&	ccd	&	I	&	Deithelm (2010)	\\
55311.3789	&	0.0022	&	40627.5	&	0.00625	&	ccd	&	II	&	Hubscher \& Monninger (2011)	\\
55311.5149	&	0.0017	&	40628.0	&	0.00664	&	ccd	&	I	&	Hubscher \& Monninger (2011)	\\
55329.8208	&	0.0004	&	40695.5	&	0.00513	&	ccd	&	II	&	Deithelm (2010)	\\
55583.8208	&	0.0012	&	41632.0	&	0.00685	&	ccd	&	I	&	Deithelm (2011)	\\
55583.9551	&	0.0003	&	41632.5	&	0.00554	&	ccd	&	II	&	Deithelm (2011)	\\
55665.7296	&	0.0008	&	41934.0	&	0.00697	&	ccd	&	I	&	Deithelm (2011)	\\
55963.9355	&	0.0007	&	43033.5	&	0.0056	&	ccd	&	II	&	Deithelm (2012)	\\
56033.7769	&	0.0019	&	43291.0	&	0.00765	&	ccd	&	I	&	Deithelm (2012)	\\
56404.6724	&	0.0001	&	44658.5	&	0.00875	&	ccd	&	II	&	Deithelm (2013)	\\
56730.4095	&	0.0018	&	45859.5	&	0.00962	&	ccd	&	II	&	Hubscher \& Lehmann (2015)	\\
56730.5433	&	0.0007	&	45860.0	&	0.00781	&	ccd	&	I	&	Hubscher \& Lehmann (2015)	\\
56737.3235	&	0.0005	&	45885.0	&	0.00749	&	ccd	&	I	&	Hubscher \& Lehmann (2015)	\\
56737.4605	&	0.0024	&	45885.5	&	0.00888	&	ccd	&	II	&	Hubscher \& Lehmann (2015)	\\
56737.5947	&	0.0007	&	45886.0	&	0.00747	&	ccd	&	I	&	Hubscher \& Lehmann (2015)	\\
57001.2206	&		&	46858.0	&	0.00675	&	ccd	&	I	&	Nagai (2015)	\\
57001.3571	&		&	46858.5	&	0.00764	&	ccd	&	II	&	Nagai (2015)	\\
57505.6933	&	0.0001	&	48718.0	&	0.00877	&	ccd	&	I	&	Samolyk (2016b)	\\
57512.7443	&	0.0005	&	48744.0	&	0.00798	&	V	&	I	&	AAVSO**	\\
57859.7726	&	0.0003	&	50023.5	&	0.00928	&	V	&	II	&	AAVSO**	\\
58159.7421	&	0.0003	&	51129.5	&	0.00862	&	V	&	II	&	AAVSO**	\\
58159.8790	&	0.0003	&	51130.0	&	0.00994	&	V	&	I	&	AAVSO**	\\
58880.7825	&	0.0002	&	53788.0	&	0.00855	&	V	&	I	&	AAVSO**	\\
58880.9184	&	0.0003	&	53788.5	&	0.00881	&	V	&	II	&	AAVSO**	\\
58942.2135	&	0.0001	&	54014.5	&	0.00805	&	ccd	&	II	&	present work	\\
58975.7097	&	0.0003	&	54138.0	&	0.00841	&	V	&	I	&	AAVSO**	\\
\hline
\end{tabular}
\end{center}
{\tiny Notes.} \tiny *We determined the times of light minima of BM UMa's light curves from SWASP database (https://wasp.cerit-sc.cz/form, see Fig. 4), **the times of minima are determined from the light curves observed by AAVSO's member (https://www.aavso.org, see Fig. 4)
\end{table}

\section{The photometric time series and their photometric elements}

\begin{figure}
\begin{center}
\includegraphics[angle=0,scale=0.4]{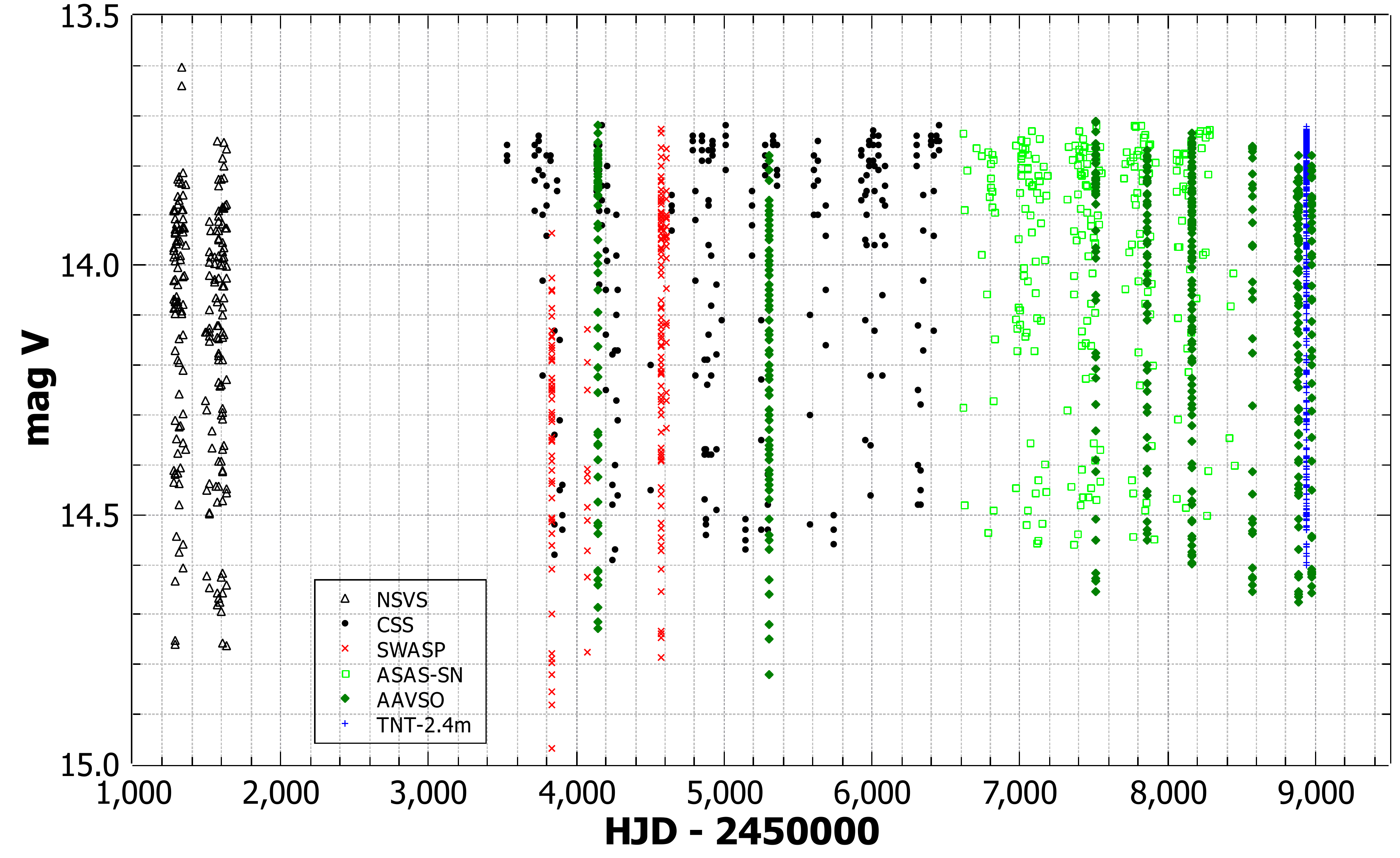}
\caption{The long-term photometric time series from all available sky-surveys and archive including data from the NSVS, CSS, SWASP, ASAS-SN and AAVSO, compared with our V-band light curve from the TNT-2.4m. The maximum brightness of light curves tends to be stable for over ten years with no any significant change or there is a very weak magnetic cycle. It indicates that the binary system may have a very weak activity or it is inactive state for decades.}
\end{center}
\end{figure}

\begin{figure}
\begin{center}
\includegraphics[angle=0,scale=0.3]{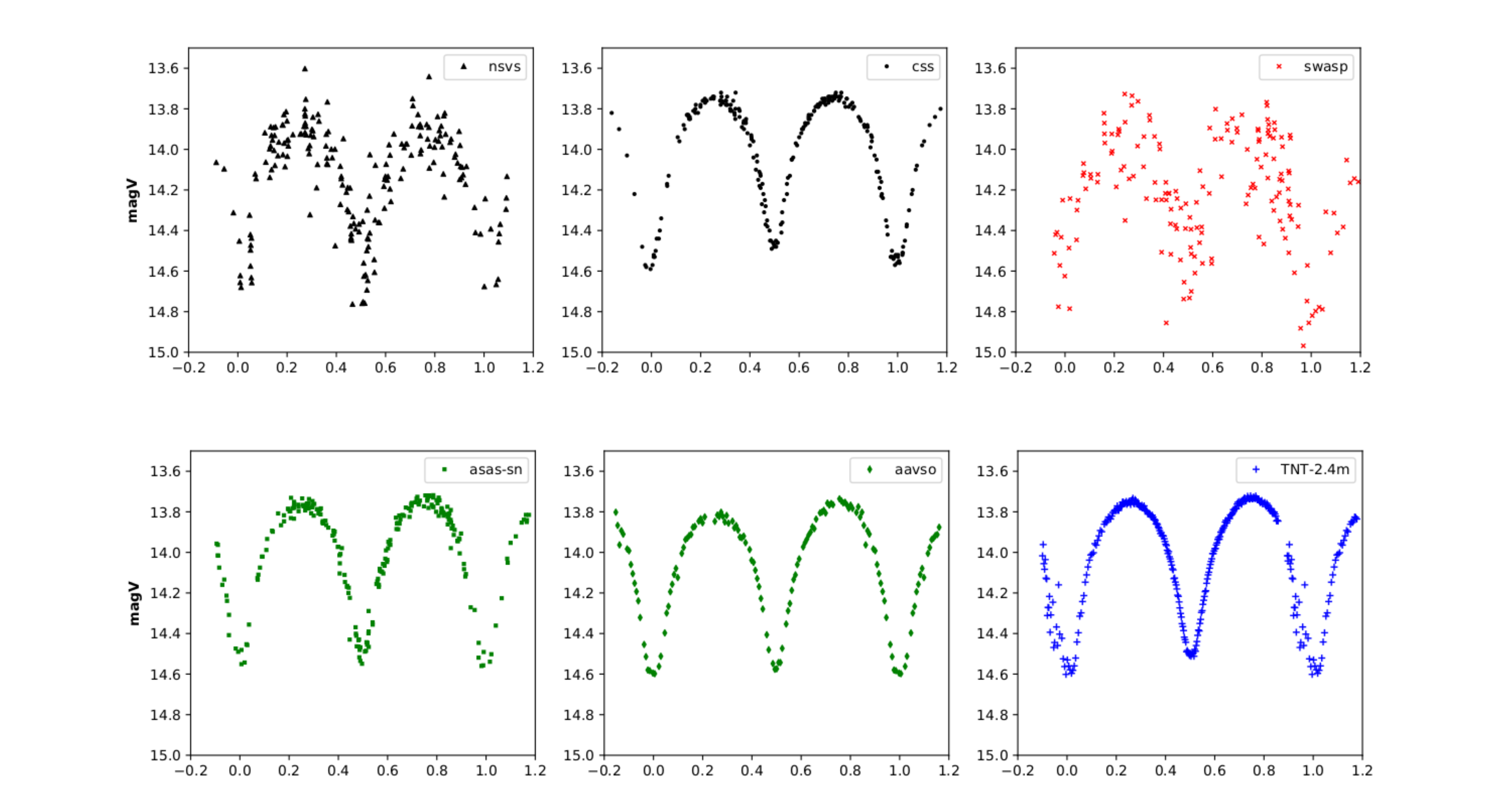}
\caption{The phased light curves from the long-term photometric time series as the same sources in Fig. 4. All are plotted separately in order to compare to each other clearly. The light curves from the CSS and ASAS-SN are similar to the V-band light curve from the TNT-2.4m with no significant differences between the maxima (light curve symmetry), while the light curve from the AAVSO (HJD 2458159.67 - 2458159.94) shows an unequal high between the maxima where the max I is lower than the max II (light curve asymmetry), compared to the one from the TNT-2.4m (HJD 2458942.05 - 2458942.31).}
\end{center}
\end{figure}

\begin{figure}
\begin{center}
\includegraphics[angle=0,scale=0.5]{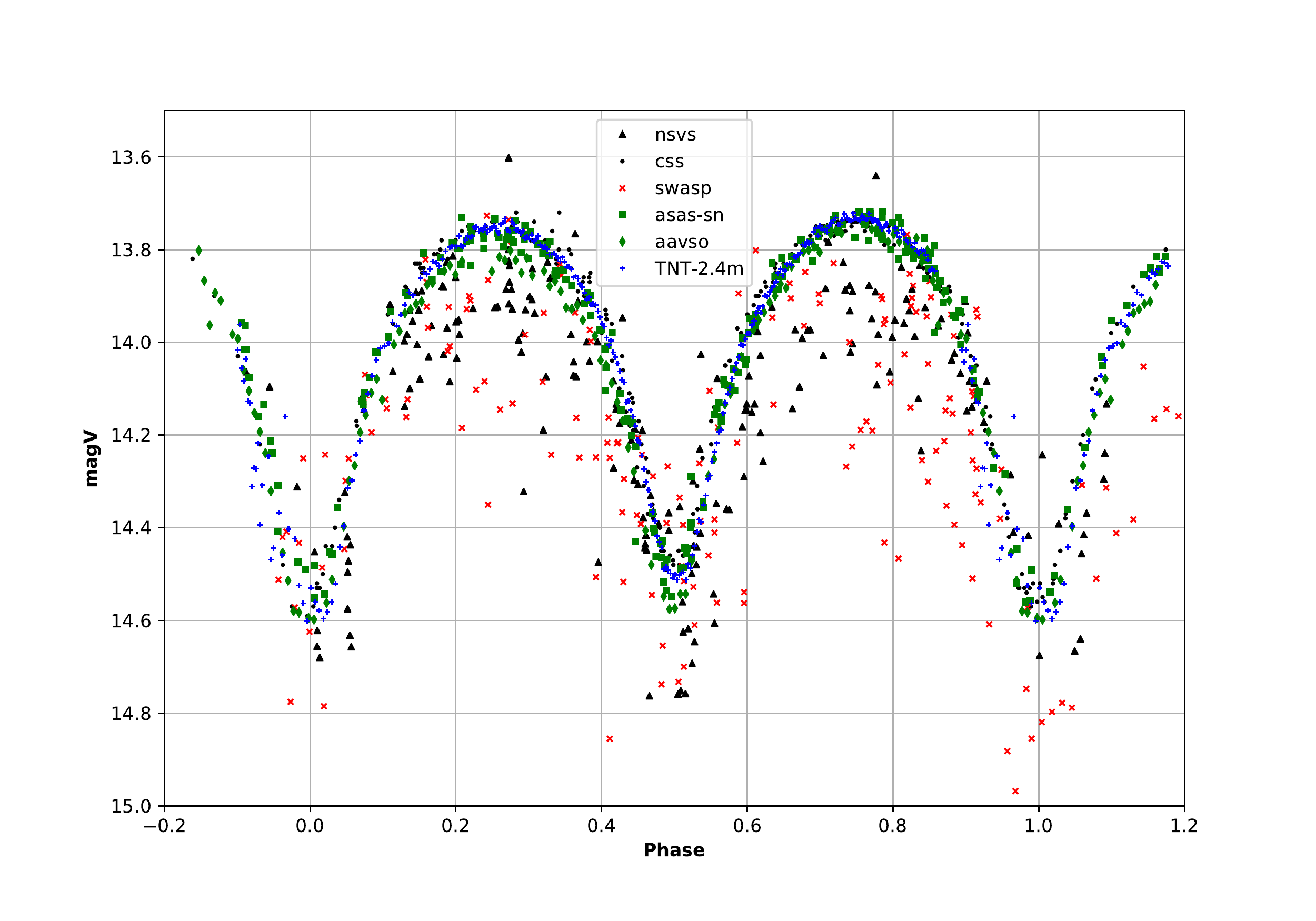}
\caption{All phased light curve from Fig 5 are put together, compared to each other and to our V-band light curve from the TNT-2.4m. It can see that all phased light curves are symmetric and well overlapped. No significant variation is found in the phased light curves, except data from the NSVS and the SuperWASP data which show a high scatter (also seen in Fig. 5).}
\end{center}
\end{figure}

To obtain the reliable parameters and examine the possible variability of BM UMa's light curves from our observations, we also compare them with the previous studies and other unpublished data e.g., sky surveys and all available online archive of light curves. The long-term photometric data are collected and plotted together as shown in Fig. 4 including data from the Northern Sky Variability Survey (NSVS; Wozniak et al. 2004), the Super Wide Angle Search for Planets (SuperWASP; Pollacco et al. 2006), the All Sky Automated Survey for Supernovae (ASAS-SN; Shappee et al. 2014), the Catalina Sky Survey's Northern (CSS; Drake et al. 2014) and the American Association of Variable Star Observers (AAVSO). We find that its long term of maximum brightness of light curves is mostly stable without a significant or dramatic change except only the light curves from SuperWASP that do not follow to the general trend formed by the other data sources. In addition, it is clear that the phased light curves of SuperWASP data in Fig. 5 and Fig. 6 show a large scatter compared to the other sources. This may because of the magnitude limit of the SuperWASP's instrument which is capable for objects having {\it V} 7-11.5 mag but BM UMa is $V$ 14 mag. Before doing the light curve analysis, the value of mass ratio $q$ has to be set first and the reliable value of $q$ should be obtained from spectroscopic observations. Unfortunately, no spectroscopic data or radial velocity (RV) curves of BM UMa were published or reported in literature (e.g. LAMOST DR6; Cui et al. 2012). However, in order to obtain reliable results, we have to search for the optimal value of photometric mass ratio $q_{ph}$ first by analyzing the completed and symmetric light curves from the CSS and the ASAS-SN databases as well as the one from our observations with the TNT-2.4m. 

For light curve modeling, we use PyWD2015 (Güzel \& Özdarcan 2020) based on the 2015 version (Wilson \& Van Hamme 2014) of the Wilson \& Devinney code (Wilson \& Devinney 1971) to investigate those light curves. To obtain the initial values for parameters in light curve modeling (e.g. the inclination ({\it i}) and so on),  the effective temperature of star 1 ($T_1$) is set as 4999 K which computed by using the equation (Collier Cameron et al. 2007) from the relation between the 2MASS {\it J-H} color index and the effective temperature for FGK dwarfs (4000 $< T_{eff} <$ 7000 K). The corresponding spectral type is estimate as K0\,V (Cox 2000) from SIMBAD database ($J$=12.929, $H$=12.428, $K$=12.309). According to the common convective envelopes (CCE) in close binary stars, the gravity-darkening coefficient ${\it g}_1 = {\it g}_2$ = 0.32 (Lucy 1967) and the bolometric albedo $A_1 = A_2$ = 0.5 (Rucinski 1969) are adopted. The selected light curves from Fig. 5 are performed separately for $q$-search as results shown in Fig. 7. The $q$-search results in Fig 7 indicate that the optimal value of $q$ is close to 2.0, thus we set $q$=2.0 as initial value for detailed search to obtain the best solution for each light curve. During the optimization, the third light ($l_3$) is also added as an adjustable parameter to check the existence of third body. The photometric solutions for $l_3$ are negative for all datasets, this may suggest that the luminosity contribution of third light is very low compared to the total light contribution from the binary system. If the presence of a third body is true, it will be a very cool dwarf. However, the existence of third body will be discussed in details on the last section. In addition, no unequal high between the primary maximum (Max-I) and the secondary maximum (Max-II) or the O'Connell effect was remarkable except the one from aavso archive. The results may suggest that it has a very weak activity and it may be inactive state for decades with no significant spot activities. Therefore, the spotted model will not be considered for our light curve analysis. With stable light curves and a very weak O'Connell effect, as well as a high inclination nearly total eclipse, reliable mass ratio can be determined via the {\it q}-search method with no needed spectroscopic RV measurements to support (see Terrell \& Wilson 2005; Li et al. 2021). The final solution for each dataset with unspotted model and photometric elements are listed in Table 3. The corresponding light curves and the geometrical structures with 3D model at phase 0.25 are also plotted in Fig 8 and 9, respectively.

\begin{figure}
\begin{center}
\includegraphics[angle=0,scale=0.6]{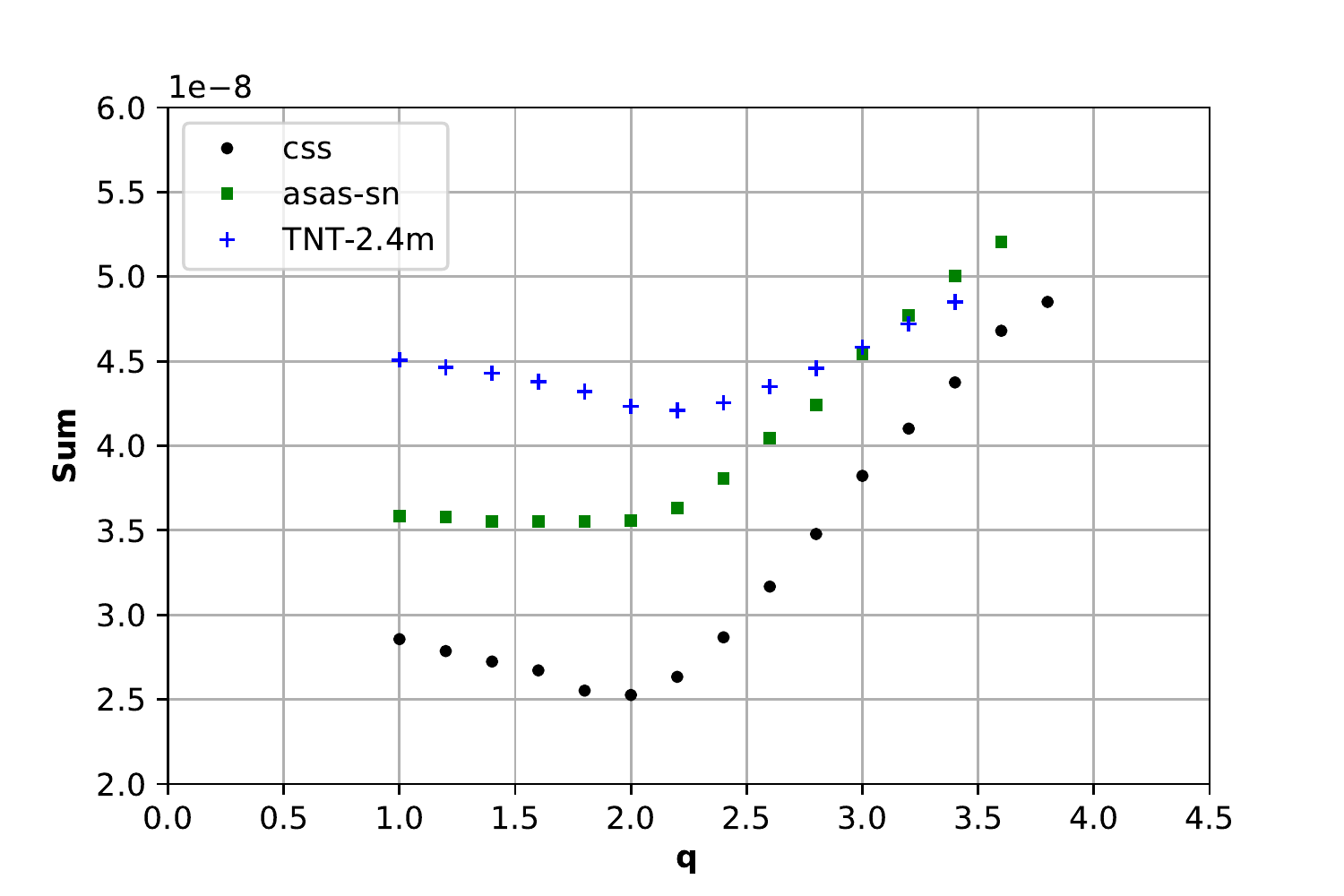}
\caption{The photometric mass ratio $q_{ph}$ from $q$-search method for the binary BM UMa. The results of $q$-search from css, asas-sn database and the one from our observations from the TNT-2.4m all suggest that the optimal value of $q$ is close to 2.0.}
\end{center}
\end{figure}

\begin{figure}
\begin{center}
\includegraphics[angle=0,scale=0.45]{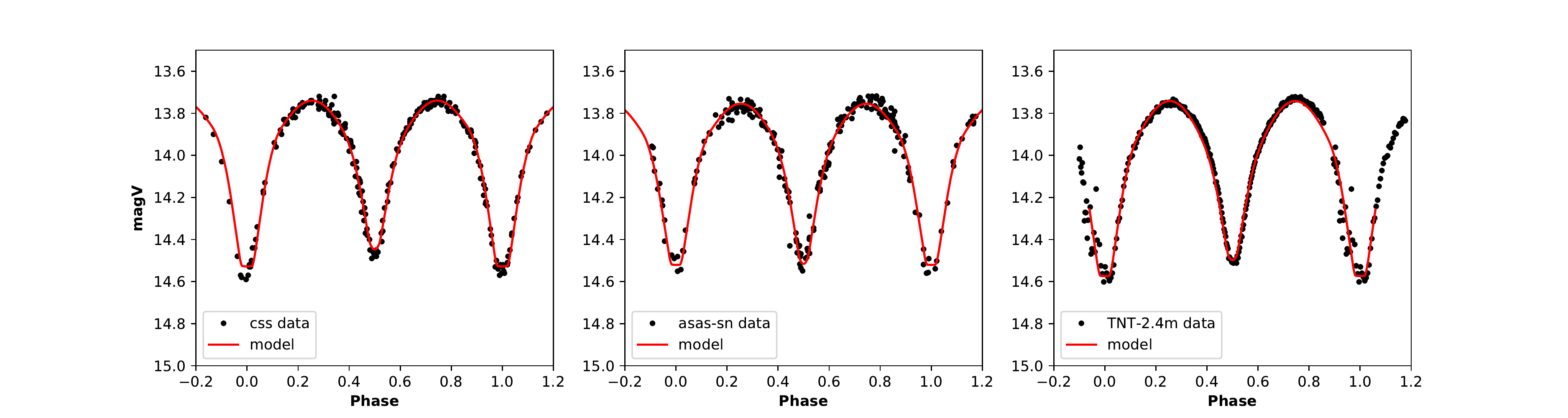}
\caption{The photometric light curve modelings for BM UMa from CSS, ASAS-SN and TNT-2.4m. The synthetic light curves (solid line) fit well with the observed light curves.}
\end{center}
\end{figure}

\begin{figure}
\begin{center}
\includegraphics[angle=0,scale=0.45]{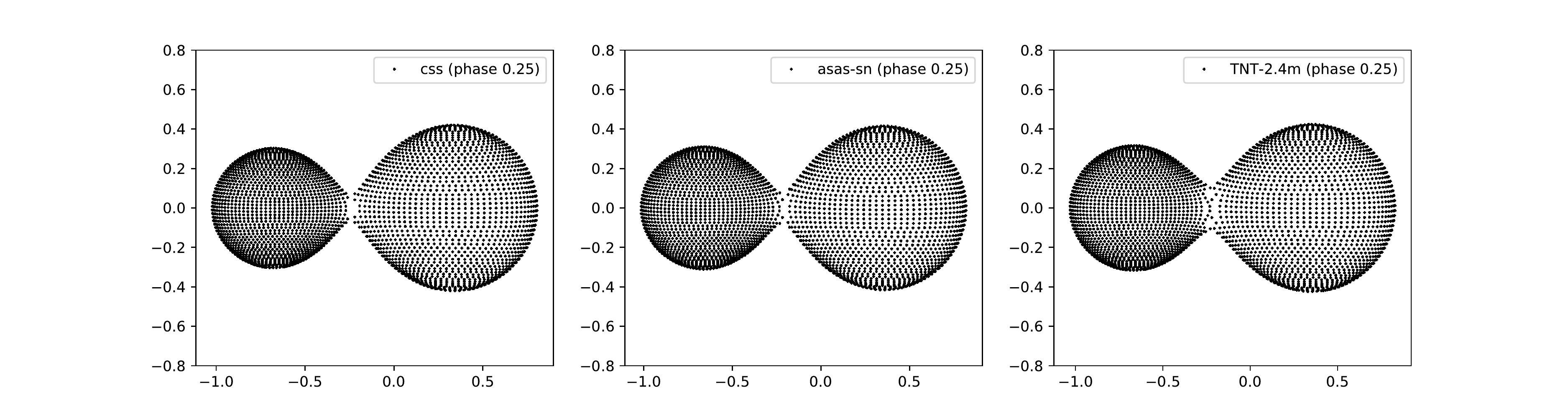}
\caption{The 3-D model of contact binary BM UMa from photometric solutions as the same sources in Fig. 8. The contact configurations show the evolution of degree of contact from 2007 to 2020 (left to right panel, respectively) with increasing fill-out factor and deeper contact state.}
\end{center}
\end{figure}

\begin{table}
\scriptsize
\caption{Photometric solutions with unspotted model and no third light}
\begin{center}
\begin{tabular}{lccccc}\hline\hline
Parameters &CSS (2007-2012) &ASAS-SN (2015-2017) &V TNT-2.4m (2020) &VRI TNT-2.4m (2020)\\
\hline
$T_1(K)$		& 4999 (fixed)	&4999 (fixed)	&4999 (fixed)	&4999 (fixed)\\
$g_1=g_2$		& 0.32 (fixed)	&0.32 (fixed)	&0.32 (fixed)	&0.32 (fixed)\\
$A_1=A_2$		& 0.50 (fixed)	&0.50 (fixed)	&0.50 (fixed)	&0.50 (fixed)\\
$q$	&2.0349($\pm0.0403$)	&1.8859($\pm0.0686$)	&1.920($\pm0.015$)	&2.0693($\pm0.0456$)\\
$1/q$	&0.4914($\pm0.0097$)	&0.5303($\pm0.0193$)	&0.521($\pm0.004$)	&0.4833($\pm0.0106$)\\
$T_2(K)$ &4783($\pm12$) &4884($\pm19$)  &4792($\pm15$)	&4726($\pm11$)\\
$i(^o)$	&88.100($\pm1.073$) &87.234($\pm1.328$)  &88.200($\pm1.463$)	&89.856($\pm0.919$)\\
$\Omega_{in}$       &2.8593 	&2.9335	&2.9157		&2.8435\\
$\Omega_{out}$      &2.5653 	&2.6188	&2.6059		&2.5539\\
$\Omega_1=\Omega_2$ &5.2483($\pm0.0558$) &5.0221($\pm0.0926$)  &5.0000($\pm0.0201$)	&5.1635($\pm0.0099$)\\
$L_1/(L_1+L_2)$(V)  &0.4067($\pm0.0001$)	&0.3928($\pm0.0001$)	&0.4197($\pm0.0055$)	&0.4269($\pm0.0001$)\\
$L_1/(L_1+L_2)$(R)  &		&	&	&0.4121($\pm0.0001$)\\
$L_1/(L_1+L_2)$(I)  &		&	&	&0.4016($\pm0.0001$)\\
$r_1(pole)$		&0.3026($\pm0.0017$)	&0.3107($\pm0.0028$)	&0.3180($\pm0.0016$)	&0.3132($\pm0.0009$)\\
$r_1(side)$  		&0.3165($\pm0.0019$)	&0.3254($\pm0.0031$)	&0.3346($\pm0.0020$)	&0.3295($\pm0.0011$)\\
$r_1(back)$ 		&0.3519($\pm0.0022$)	&0.3621($\pm0.0036$)	&0.3787($\pm0.0035$)	&0.3742($\pm0.0019$)\\
$r_2(pole)$ 		&0.4209($\pm0.0062$)	&0.4162($\pm0.0112$)	&0.4263($\pm0.0023$)	&0.4329($\pm0.0009$)\\
$r_2(side)$ 		&0.4484($\pm0.0083$)	&0.4429($\pm0.0148$)	&0.4559($\pm0.0031$)	&0.4639($\pm0.0012$)\\
$r_2(back)$   	&0.4783($\pm0.0116$)	&0.4739($\pm0.0212$)	&0.4905($\pm0.0045$)	&0.4985($\pm0.0016$)\\
$f(\%)$	&8.8($\pm 4.3$)	&11.3($\pm 5.6$)	&23.2($\pm 3.4$)	&31.0($\pm 1.7$)\\
$\Sigma{W(O-C)^2}$ $(10^{-8})$	&2.31	&3.25	&3.57	&4.20\\
\hline
\end{tabular}
\end{center}
\end{table}

\section{The study of W-type contact binaries (for 0.22 $<$ P $<$ 0.3 days)}

According to the recent studies on orbital period distribution in EW-type binaries e.g., Rucinski (2007); Jiang et al. (2014); Qian et al. (2017), it is noticed that the period distribution is between 0.27 to 0.29 days and most of surface temperatures of those systems correspond to solar type star (e.g. G to K spectral-type stars). In addition, it is found that most of short period (P $<$ 0.3 days) contact binaries are W-type contact systems while A-type systems are composed of A to F spectral-type stars with longer orbital period (e.g. P $>$ 0.3 days). On the contrary, many studies show that there are both A-type and W-type contact binaries that have orbital period below the period limit P $<$ 0.22 days (Dimitrov \& Kjurkchieva 2015; Jiang et al. 2015), but their origin and evolution, as well as their connection are poorly understood and investigated. The formation and evolution of contact systems can be divided into different 3 groups e.g., period below 0.22 days, period between 0.22 to 0.3 days and period longer than 0.3 days. Based on preliminary results, we will show that those W-type contact binaries which have periods longer than 0.3 days show a high rate of period increase. This may suggest that its formation and evolution may differ from other groups. For detailed and further investigation will be reported in the next publication. For contact binaries with periods shorter than 0.22 days are commonly composed of late K to M spectral-type stars which are low luminosities and difficult to monitor with small telescopes for both long-term and short-time with high cadence photometries. Nowadays, many contact binaries with periods below the period limit are found and studied but not in detailed because of some limitations (i.e., it is required to monitor their long-term period variations and investigations of their long-term physical parameters changes for each binary system such we investigated BM UMa). Thus, this study we will focus only on W-type contact systems which have short periods 0.22 $<$ P $<$ 0.3 days. 

In order to understand the origin and evolutionary path of W-type contact binaries and to find a possible evolutionary correlation between W-type and A-type contact systems, we compiled all physical parameters and observed period changes from well-investigated W-type contact binaries in the literature as listed in Table 4. Based on those parameters from many investigators, we plot the mass ratio $q$ and $1/q$ against the orbital period (days) and the fill-out factor f (\%) as shown in Fig. 10. The figure shows that the period distribution is in the range between 0.25 to 0.30 days which is in agreement to the previous results from many authors above. According to Fig 10, it is also noticed that the most population of short period W-type contact systems is located at mass ratio $q$ = 2.0 (or $1/q$ = 0.5). It is also found that there is an existence of maximum mass ratio $q$ close to 10 (or $1/q$ close to 0.1) with high fill-out factor (f $>$ 40\,\%) and long period (P $>$ 0.3 days), but no W-type system which has fill-out factor larger than 65\,\% is found. In addition, no shallow contact binary with mass ratio $q$ = 10 or $1/q = 0.1$ or $1/q < 0.1$ is found. This may imply that no W-type system can exist or keep its W-type configuration in this evolutionary stage.

To investigate more details, we also plot more diagrams with the same parameters as Fig 10 but against the period change rate (dP/dt) either positive (increasing trend) or negative (decreasing trend) change as shown in Fig. 11. Both two plots in above panel of Fig. 11 have shown that the orbital period change rate is rapidly increased (positive trend) with high rate more than 300 $\times 10^{-8}$ days/year when $q$ close to 10 (or $1/q$ close to 0.1). Based on Fig 11, we fit the data by using a least-squares method and the fitting results are the upward parabolic curves for both data sets, yielding the equation for the correlation between $q$ and dP/dt (F test = 52.53) as displayed in Eq. 3, while the correlation between $1/q$ and dP/dt (F test = 9.32) is shown in Eq. 4, respectively. 

\begin{equation}
dP/dt = 82.48 (\pm 16.60) - 46.49 (\pm 9.68) \times q + 7.39 (\pm 1.02) \times q^{2}, 
\end{equation}

\begin{equation}
dP/dt = 192.60 (\pm 38.85) - 648.94 (\pm 150.63) \times q^{-1} + 549.55 (\pm 132.55) \times q^{-2}
\end{equation}

From the Eq. 3, it suggests that the lowest change rate occurs when $q$ close to 3.14 and the derived minimum period change rate is 9.36 $\times 10^{-8}$ days/year which can be interpreted as a period increase or period decrease because it just indicates the direction of mass transfer i.e., from the less massive component to the more massive one or from the more massive to the less massive, respectively. On the other hand, Eq. 4 gives a different result compared to Eq. 3 i.e., its optimize value derived as $1/q$ = 0.59 (or $q$ = 1.69) for the minimal dP/dt where its lower limit of period change rate is derived as 1.02 $\times 10^{-8}$ days/year. As a result from Fig. 10, from Eq. 3 and Eq. 4, those may suggest that dP/dt has a relationship with its mass ratio $q$ rather than $1/q$. Moreover, one can see that $1/q$ cannot tell us about period distributions of W-type contact systems, while the parameter $q$ shows clearly that the most population of W-type systems is located at $q \sim 2.0$ with fill-out factor covering from $0\,\% < f < 25\,\%$ and the range of period distribution is 0.25 to 0.30 days. The analysis results of $q$ are in agreement with the observations. Therefore, the mass ratio $q$ is a crucial parameter that can predict the evolutionary path of the W-type contact systems (where 0.22 $<$ P $<$ 0.3 days). For another cases, e.g. some W-type systems show a long-term period increase or an oscillation between the marginal and shallow phase which explained by thermal relaxation oscillation idea (TRO: Lucy 1976; Flannery 1976) will be discussed in details in the next publication.

Based on the results above, the parameter $q$ also shows that at the early evolution of W-type contact binaries when $q$ close to 10 the binaries have a high period change rate. This may because of mass transfer and the high mass differences between the less massive and the more massive components in the binaries. When $q$ is gradually decreased close to $q$ = 3.14 (i.e., due to mass transfer or AML) near the late W-type contact evolution, the period changes undergo a slow rate with dP/dt = 9.36 $\times 10^{-8}$ days/year. The binaries will evolve into A-type contact systems after $q$ decreased close to 1.0 at the end stage of evolution of W-type configuration. This is one possible scenario of evolutionary track of W-type systems by changing from W-type into A-type contact binaries. This assumption can answer the question that why we cannot find the W-type systems when the fill-out factors of the binaries are larger than 40\,\% whereas the binaries undergo the long-term period decrease. In the other word, when their fill-out factors reach the value of 40\,\% as well as their mass ratios $q$ close to 1.0, those W-type contact binaries cannot keep W-type configurations while experiencing secular period decrease due to mass transfer (the mass transfer still takes place as the same direction) or plus AML. In fact, when degree of contact reaches the maximum value (e.g. f = 40\,\%) the contact configuration cannot be broken and changed to be near contact or EA-type eclipsing binaries because of secular period decrease and under the AML-controlled stage of the evolutionary scheme. Therefore, when the values of mass ratio $q$ below 1.0, those binaries will change into A-type contact configuration. For the case that W-type systems show a high fill-out factor larger than 40\,\% as displayed in Fig. 10 and 11, suggesting that the W-type systems with long-term period increase (mass transfer from the less massive component to the more massive one or $q < 1.0$) and high fill-out factor are at the early age of W-type evolution or may be a misclassification between A-type or W-type contact systems. This may need statistical study with more targets and comprehensive investigations from literature to confirm this assumption and its scenario. 

\begin{table*}
\scriptsize
\caption{Parameters of W-type contact binaries which have orbital periods shorter than 0.3 days}
\begin{tabular}{lllllrrcccl}\hline\hline
Star &Sp. &Period &$q$ &$1/q$ & $f$ &$\mathrm{d}P/\mathrm{d}t$ &Cyclic &$l_3$ &LTTE &Ref.\\
 & & (days)& & &($\%$)&($10^-{8}$ d/y)& &($\%$) &$M_3$($M_{\odot}$)&\\
\hline
KIC 9532219&G9&0.1981&1.20 &0.833&-&-52.7&yes&76&0.09&Lee et al. (2016)\\
NSVS 7179685&K8 &0.2097 &2.13 &0.469 &19.3 &-&-&1.2&-&Dimitrov et al. (2015)\\
SWASPJ015100&K4 &0.2145 &3.13&0.319 &14.6&-&-&no&-&Qian et al. (2015)\\
CC Com&K6 &0.2207&1.90 &0.527&16.7&-2.0&yes&no&0.07&Yang et al. (2009a)\\
V1104 Her &K7 &0.2279 &1.60 &0.625 &15 &-2.9 &yes &no &m3+m4 &Liu et al. (2015)\\
1SWASPJ161335 &K0 &0.2298 &1.10 &0.909 &19 &-42.6 &yes &no &0.15 &Fang et al. (2019)\\
V523 Cas &K4 &0.2337 &1.76 &0.518 &21.6 &+34.4 &yes &no &0.60 &Jeong et al. (2010), (Castelaz 2014)\\
YZ Phe &K2 &0.2347 &2.63 &0.379 &10 &-2.6 &yes &2 &0.13 &Sarotsakulchai et al. (2019b)\\
RW Com&K2&0.2373 &2.10 &0.471 &6 &0&yes&no &m3+m4 &Ozavci et al. (2020)\\
FY Boo &K3 &0.2412 &2.55 &0.392 &11 &0 &yes &no &0.16 &Samec et al. (2011)\\
1SWASPJ064501&K5 &0.2486 &2.11 &0.474 &15.3 &- &- &2&-&Liu et al. (2014a)\\
BI Vul&K3&0.2518&1.04 &0.964&8.7&-9.5&yes&no&0.30&Qian et al. (2013)\\
EI CVn &K5 &0.2608 &2.17 &0.461 &21.0 &-31.1 &unclear&no &- &Yang (2011)\\
PZ UMa&G7 &0.2627 &5.62&0.178&38.8&-&yes&no&0.84 &Zhou $\&$ Soonthornthum (2019a)\\
V1197 Her&K1 &0.2627 &2.61&0.383 &15.7&-25.8 &unclear &no &- &Zhou $\&$ Soonthornthum (2020)\\
EH CVn &G9 &0.2636 &3.33&0.300 &19.2 &-5.2 &unclear &no &- &Xia et al. (2018)\\
GV Leo &K2 &0.2667 &5.32 &0.188 &17.7 &-49.5 &unclear&2&- & Kriwattanawong $\&$ Poojon (2013)\\
V336 TrA&K1&0.2668&1.39 &0.716&15.7&- &- &no&-&Kriwattanawong et al. (2018)\\
CSTAR 038663&K3&0.2671&1.12 &0.890&10.6&-&yes&$<1$&0.63&Qian et al. (2014)\\
IL Cnc&K3&0.2676&1.76&0.568 &8.9 &-0.68 &yes &no &0.06 &Liu et al. (2020)\\
DD Com &G5 &0.2692 &3.69 &0.271 &8.8 &+0.01 &unclear &no &-&Zhu et al. (2010)\\
FG Sct &K4 &0.2706 &1.35 &0.740 &21.4 &-6.39 &unclear&no&- &Yue et al. (2019)\\
BM UMa&K0&0.2712&2.00 &0.500&23.2&-3.36&unclear&no&- &this study\\
EF CVn &G5 &0.2720 &3.49 &0.286 &32.0&+45.2 &yes&no&0.22& Xia et al. (2018)\\
VW Cep &K1 &0.2783 &3.31 &0.302 & &-16.9 &yes &no &m3+m4 &Mitnyan et al. (2018)\\
V1005 Her &G5 &0.2789 &3.29 &0.303 &11.2 &-15.9 &yes &4 &0.45 &Zhu et al. (2019)\\
BX Peg &G5&0.2804 &2.69 &0.372&23.1 &-9.84&yes&no &m3+m4 &Lee et al. (2009)\\
AQ Com &G8 &0.2813 &2.86 &0.350 &21.4 &- &yes &7-11 &0.08 &Liu et al. (2014c)\\
V357 Her &K2 &0.2820 &3.57 &0.280 &11 &- &- &no &- &Branicki \& Pigulski (2002)\\
AD Cnc &K0 &0.2827 &1.30 &0.770 &8.3 &+49.4 &yes &no &m3+m4 & Qian et al. (2007)\\
GSC 2765-0348&G4&0.2835&3.19 &0.313&34.0&-&unclear &no &-&Samec et al. (2012)\\
V524 Mon &G5&0.2836&2.10 &0.476&7.7 &-.015 &yes&no&0.26&He et al. (2012)\\
XY Leo &K2 &0.2841 &1.64&0.609 &6.7&+2.7 &yes&8&0.98&Yakut et al. (2003)\\
RW Dor &G4&0.2854&1.59 &0.630&11&-1.4&unclear&no&0.09&Sarotsakulchai et al. (2019a)\\
ER Cep &G8 &0.2857 &2.22 &0.450 &62.0 &+9.7 &yes &no &m3+m4 &Liu et al. (2011)\\
V2790 Ori &G1 &0.2878 &2.93 &0.341 &20.9 &+10.3 &unclear &no &- &Kriwattanawong et al. (2019) \\
V842 Cep &K0 &0.2889 &2.28 &0.438 &8.7 &-15 &unclear &no &- &Li et al. (2021)\\
EP Cep&K1 &0.2897 &6.51 &0.154 &10.7 &-37.3 &unclear &no&-&Zhu et al. (2014)\\
PS Vir &G2 &0.2898 &3.4 &0.305 &14.4 &unclear&unclear&0.3 &0.12 &Yuan et al. (2018)\\
LO Com&K0&0.2864&2.48 &0.404&3.2&-11.8&yes&no&-&Zhang et al. (2016)\\
V1799 Ori &K1 &0.2903 &1.33 &0.749 &3.5 &+1.8 &unclear &no &- &Liu et al. (2014b)\\
GSC 03526 &K2&0.2922&2.84 &0.351&18.2&-&yes&no&0.57&Liao et al. (2012)\\
UW CVn &G5 &0.2925 &4.08 &0.245 &20 &unclear &unclear &no &- &Kopacki $\&$ Pigulski (1995)\\
V354 UMa&G1&0.2938&3.62&0.276&10.7&unclear&unclear&no &-&Michel et al. (2019)\\
V873 Per &K0 &0.2949 &2.50 &0.399&18.1&-33.1 &yes&no&0.20&Kriwattanawong $\&$ Poojon (2015)\\
OU Ser &F9 &0.2968 &5.77 &0.173 &30.7 &- &- &- &- &Pribulla \& Vanko (2002)\\
BN Ari&K1&0.2994&2.55&0.392&6.9&-47.2&yes&0.2&0.23 &Alton et al. (2018b)\\
IK Boo&G2&0.3031&1.14 &0.873&2.2&-21.7&yes&-&0.21&Kriwattanawong et al. (2017)\\
CE Leo &K0 &0.3034 &1.88 &0.533 &15.8 &+16.3 &yes &no &0.33 &Yang et al. (2013)\\
V2284 Cyg&G7&0.3069& &0.345&39.2 &-29.7 &yes&no&0.04&Wang et al. (2017)\\
EQ Cep &K1 &0.3070 &1.90 &0.526 &62.1 &+117 &unclear &no &- &Liu et al. (2011)\\
V1191 Cyg & &0.3134 &9.36 &0.107 &57.9 &+313 &unclear &no &- &Ostadnezhad et al. (2014)\\
FV CVn &G6 &0.3154 &1.07 &0.930 &4.6 &-113&yes&2.5 &- & Michel et al. (2019)\\
V0599 Aur &G5 &0.3165 &1.61 &0.617 &4.3 &-30.7 &yes &no &0.27 &Hu et al. (2020)\\
TY Boo &G3&0.3171 &2.15 &0.465&7.6 &-3.6 &yes &no &0.52 &Christopoulou et al. (2012)\\
\hline
\end{tabular}
\end{table*}

\begin{figure}
\begin{center}
\includegraphics[angle=0,scale=0.7]{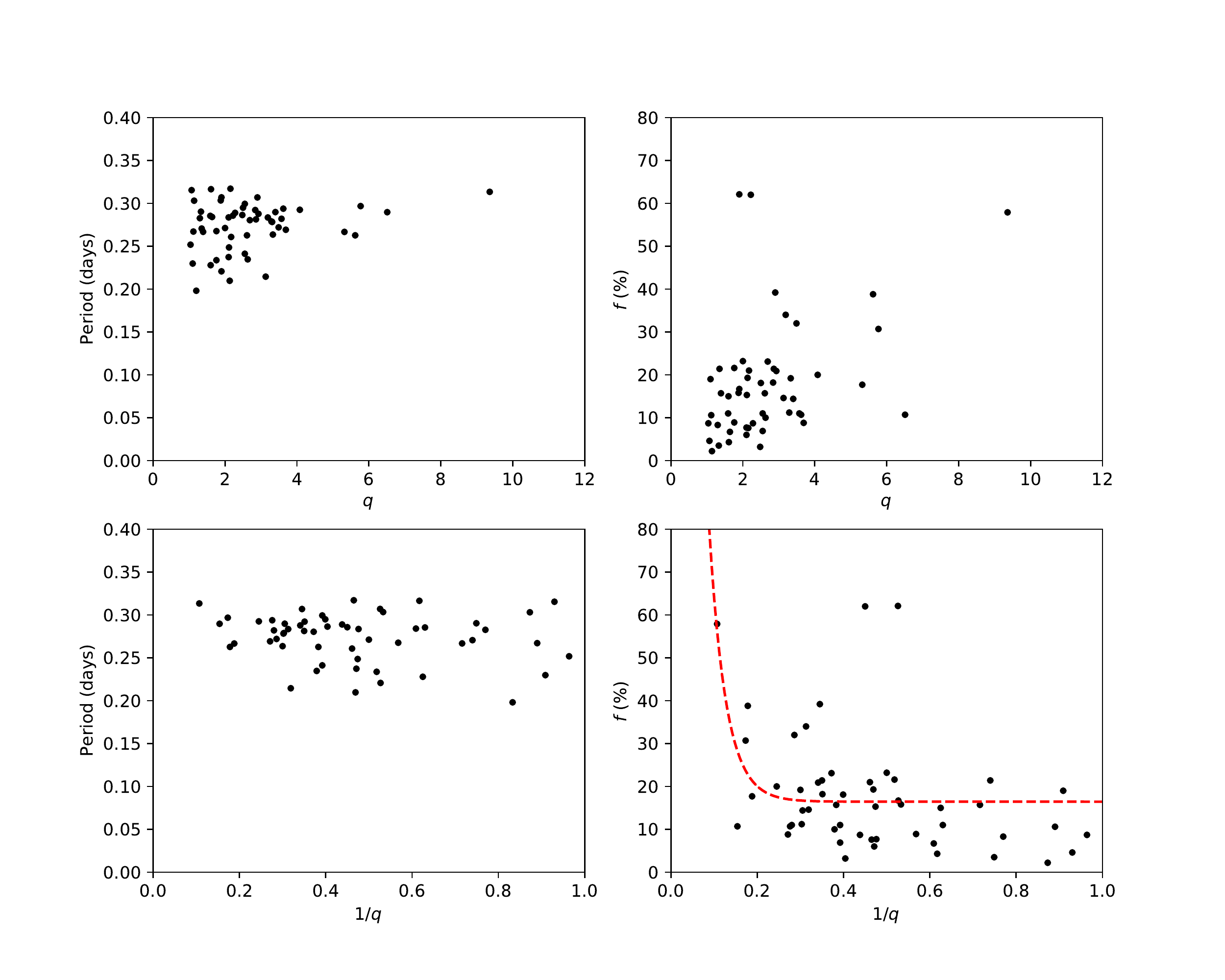}
\caption{The plots of mass ratio $q$ and $1/q$ against the orbital periods and the fill-out factors. The results show that the period distribution of W-type contact systems is in the range between 0.25 to 0.30 days with population of mass ratios between 1 to 6. The figure also shows that the most population of short period W-type systems is located at mass ratio $q$=2.0 (or $1/q$=0.5) covering fill-out factor $0\,\% < f < 25\,\%$. There is an existence of maximum mass ratio $q$ close to 10 (or $1/q$ close to 0.1) with high fill-out factor (f $<$ 60\,\%) and long period (P $>$ 0.3 days). No $q$ = 10 or $1/q$ less than 0.1 is found from the present study.}
\end{center}
\end{figure}

\begin{figure}
\begin{center}
\includegraphics[angle=0,scale=0.7]{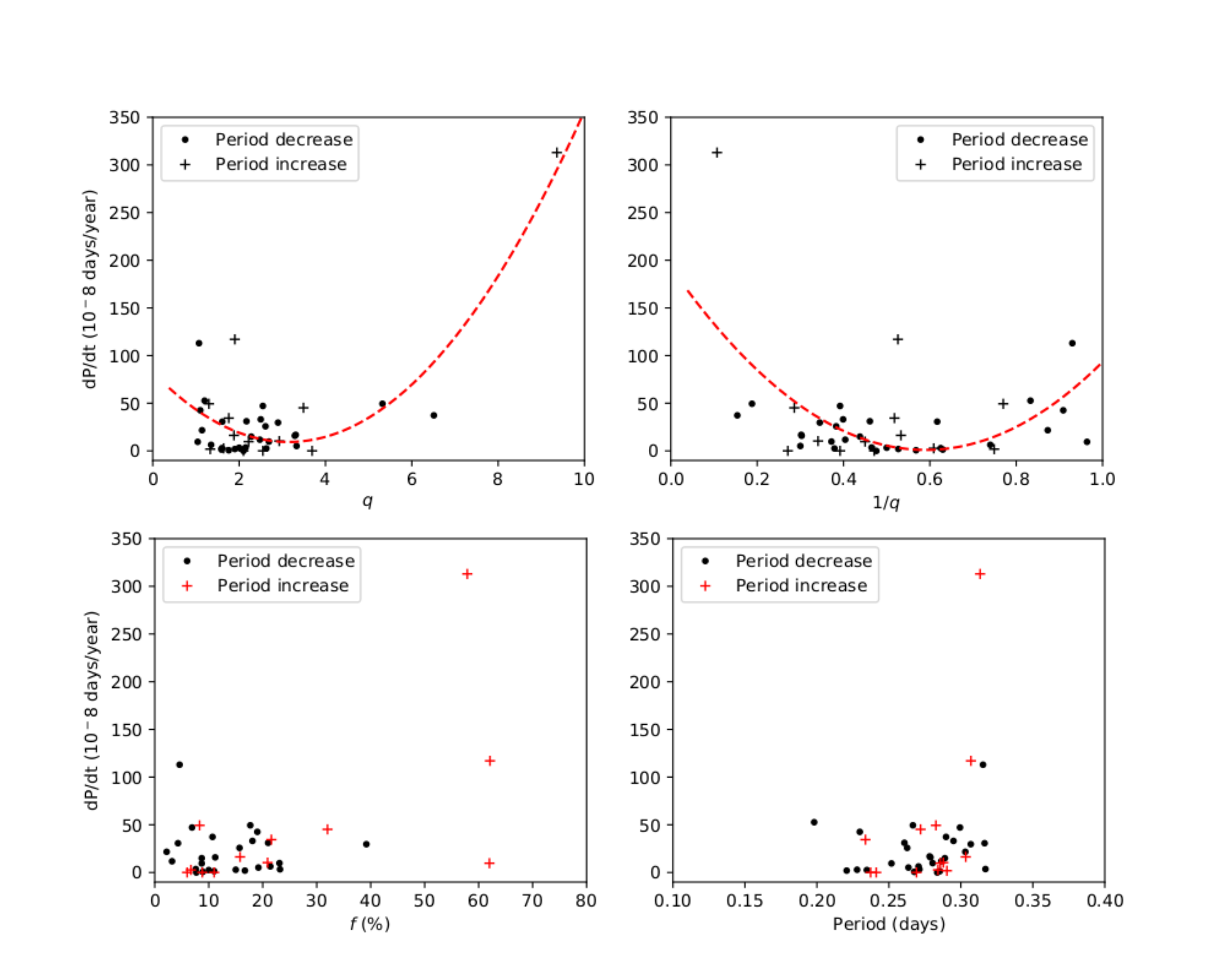}
\caption{The obove panels show the plots of mass ratio $q$ and $1/q$ against the period change rate dP/dt. The dashed lines are the curves fitting to the data with increasing trend for both $q$ and $1/q$. The bottom panels are the plots for fill-out factors (f) and period distributions against the period change rate. The results show that no W-type systems with period decrease that have f $>$ 40\,\% are found, except the three W-type systems with period increase that show high f (f $>$ 40\,\%) and high period change rates. The figure also shows that where the period P near 0.31 days, the period change rate is increasing rapidly.}
\end{center}
\end{figure}

\section{Discussions and conclusions}

BM UMa has been studied by many authors (e.g., Samec et al. 1995; Yang et al. 2009, Virnina et al. 2010), but  poorly investigated in details and neglected for over ten years since 2010. BM UMa is a short period 0.2712\,d with W-subtype contact binary (the hotter companion is a smaller one) where both companions are early K-type dwarfs. According to our study of long-term period variations, we found that the period is decreasing with a rate of $\mathrm{d}P/\mathrm{d}t = -3.36(\pm 0.02)\times10^{-8}$ d $\textrm{yr}^{-1}$. In general, the long-term period decrease can be explained by mass transfer from the more massive component to the less massive one or by the angular momentum loss (AML) via magnetic braking, or by the combination of both mechanisms. To check this, we assume the mass of hotter component by estimation from Cox (2000) where the mass ratio ($q$ = $m_2$/$m_1$) is about 2.0 (from photometric solution in Table 3). Thus, with $T_{eff}$ = 5000 K (Sp. = K0\,V), the mass of the hotter component (the less massive $m_1$) $\sim$ 0.39$M_{\odot}$, while the cooler one (the more massive $m_2$) $\sim$ 0.79$M_{\odot}$ and the total mass of the system $M$ $\sim$ 1.18$M_{\odot}$. The estimated absolute parameters for BM UMa are; $M_1=0.39M_{\odot}, M_2=0.79M_{\odot}, a=1.86R_{\odot}, R_1=0.64R_{\odot}, R_2=0.85R{\odot}, L_1=0.229L_{\odot}, L_2=0.343L_{\odot}$, respectively.

If the long-term period decrease is due to conservative mass transfer from the more massive component to the less massive one, the mass transfer rate can be determined with the following equation (Kwee 1958),

\begin{equation}
\frac{\dot{P}}{P} = -3\dot{M_2}(\frac{1}{M_1} - \frac{1}{M_2})
\end{equation}

The result is $\mathrm{d}M_{2}/\mathrm{d}t = 3.26\times10^{-8}$ $M_{\odot} \textrm{yr}^{-1}$. The timescale from computed mass transfer rate can be estimated as $M_2/\dot{M_2} \sim 2.42\times10^7 \textrm{yrs}$ or 24.2 Myr but the timescale from observed period decrease is $P/(\mathrm{d}P/\mathrm{d}t) \sim 8.07\times10^6 \textrm{yrs}$ or 8 Myr which is the lower limit, while the timescale by thermodynamic processes is 67.7 Myr which is the upper limit. In addition, we also examine the period decrease by the angular momentum loss (AML) via magnetic stellar wind from the well-known equation given by Bradstreet \& Guinan (1994). 

\begin{equation}
\dot{P} \approx 1.1\times10^{-8}q^{-1}(1+q)^2(M_1 + M_2)^{-5/3}k^2
         \times(M_{1}R^{4}_{1} + M_{2}R^{4}_{2})P^{-7/3},
\end{equation}
where $k^2$ is the gyration constant ranging from 0.07 to 0.15 for solar type stars. By adopting a value of $k^2$ = 0.1 (Bradstreet \& Guinan 1994), the rate of period decrease due to spin-orbit coupled AML can be derived as $\mathrm{d}P/\mathrm{d}t = -2.75\times10^{-8}$ d $\textrm{yr}^{-1}$ which is close to the present observed period change rate. However, based on long-term photometric time series it can see that the maximum brightness is stable for decades without significant variations (e.g., flares, outburts or magnetic cycle activities), except a very weak spot activity which had occurred in short life (see Samec et al. 1995) and recent unpublished data from AAVSO in Fig 5. This can be interpreted that the activity of BM UMa is quite weak but the period decrease from AML cannot be neglected for this case. Therefore, the plausible explanation for long-term period decrease in BM UMa is the mass transfer or the combination of the two mechanisms (mass transfer and AML). In addition, since the short-term variation or periodic change in the $O - C$ curve is not clear, as well as no detected third light, the third body will not be discussed further details here.

According to the results of our statistical study on W-type contact binaries, it is found that the most population of the binaries is located at $q$ = 2.0 with period distribution $0.25 < P < 0.3$ days and fill-out factor $0 < f < 25\,\%$. In addition, it is also found that the mass ratio $q$ ($> 1$) is likely a main parameter, as well as the period change rate, to control the direction and evolutionary stages of W-type systems. There are strong evidences to support that there is evolutionary correlation between W-type and A-type contact systems by transforming from late W-type systems into A-type systems when the value $q$ decreased and close to 1.0 while the orbital period is decreasing by mass transfer and AML. There are no W-type systems with $q$ = 1 or $q < 1$, as well as the fill-out factor $f > 40\,\%$ to be found because they have already evolved into A-type contact systems. For the cases that W-type systems with high $q$ (close to 10), high fill-out factor and high period change rate, they can be considered as early W-type systems at the beginning of the evolutionary stage or they are at the transitional phase and oscillated between marginal and shallow stage predicted by TRO theory. For the case of W-type systems with orbital period P $>$ 0.3 days (e.g., periods close to 0.31 days), they show a high period change rate. This is not clear about the cause behind, this will be investigated and reported in details for the next publication. However, these results indicate that there is a different path of evolution among W-type and A-type, which is in agreement to many studies (e.g., Zhang et al. 2020; Li et al. 2021). 

Based on the observations and analysis results e.g., growing fill-out factor and continuing period decrease, as well as the results of statistical study of W-type contact binaries, those may suggest that BM UMa is at the pre-transition stage from the late W-type into A-type contact as seen a slow rate of period change near the minimum. Both variations of fill-out factor and period have been found in many systems e.g., CC Com (Yang et al. 2009a) and V599 Aur (Hu et al. 2020). More examples e.g., V873 Per shows clearly about increasing fill-out factor from 5.5\,\% (Samec et al. 2009) to 18.1\,\% (Kriwattanawong \& Poojon 2015), V502 Oph e.g., $f$ = 22\,\% (Deb \& Singh 2011) to $f$ = 35.3\,\% (Zhou et al. 2016). After the evolution of BM UMa reaches the stage where mass ratio $q$ close to 1 and goes further the point $q < 1.0$, the binary will evolve into A-subtype as a deeper normal overcontact binary where the mass transfer takes place from the less massive component ($m_2$) to the more massive one ($m_1$), similar to the present evolutionary stage of V802 Aql (P = 0.2677\,d; Yang et al. 2008) where its orbital period is increasing and the fill-out factor also gradually grows ($f$ = 35\,\%) at the beginning stage of A-subtype before becoming a deep low-mass ratio overcontact system. By this way, the evolution of BM UMa will end as a merger or rapid-rotating single star when the mass ratio meet the critical value ($q < 0.094$, see Rasio 1995; Arbutina 2007; Zhu et al. 2016) and the binary will produce a red nova similar to the case of the red outburst V1309 Sco (Tylenda et al. 2011) and the recent study of KIC 9832227 (Gazeas et al. 2021). The changing of contact type from W-type to A-type has been found in many systems e.g., EM Psc (Yang et al. 2005; Qian et al. 2008). The study of binary system such BM UMa will be an important observational evidence to reveal the evolutionary correlation between W-type and A-type contact binary. The more observations and investigations of W-type systems will be very useful to understand their formation and evolution, as well as a proof of the assumption about the link between W-type and A-type contact systems.

\bigskip

\vskip 0.3in \noindent

The authors are very grateful to the referee for very helpful comments and suggestions to improve the quality of the manuscript. This work is supported by Thailand Science Research and Innovation (TSRI): Fundamental Fund (Grant No. 88309). We would like to thank NARIT's Time Allocation Committee (TAC) to manage and schedule time for us to use the TNT-2.4m for the observations. Based on observations made with the TNT focal reducer and the ARC 4K camera at the Thai National Observatory (TNO) under the program ID 1, which is operated by the National Astronomical Research Institute of Thailand (Public Organization). Special thanks to Mr Sawang Kangkriangkrai and TNO's staff for support during the observing run.
This research has made use of the VizieR catalog access tool, CDS, Strasbourg, France. This work was also made use of data from the AAVSO international database (https://www.aavso.org/). This paper also makes use of data from the DR1 of the WASP data (Butters et al. 2010) as provided by the WASP consortium, and the computing and storage facilities at the CERIT Scientific Cloud, reg. no. CZ.1.05/3.2.00/08.0144 which is operated by Masaryk University, Czech Republic. 

\section{Data availability}
The data underlying this article are available in the article and in its online supplementary material or upon request to the corresponding author.

\end{document}